\documentclass[prb,10pt,floatfix,amsmath,twocolumn,superscriptaddress]{revtex4}
\usepackage[ansinew]{inputenc}
\usepackage{graphicx}
\usepackage{pst-all}
\usepackage{color}
\usepackage{dcolumn}
\usepackage{amsmath}
\usepackage{amsthm}
\usepackage{bm}
\usepackage{layout}
\usepackage{float}
\usepackage{txfonts}
\usepackage{amsfonts}
\usepackage{amssymb}%
\usepackage[ansinew]{inputenc}
\usepackage{array}
\usepackage{multirow,bigdelim}
\usepackage{slashbox}
\usepackage{enumitem}

\begin{document}
\title{Three-component bosons in TiS, ZrSe and HfTe}

\author{Sami Ullah}
\affiliation{Shenyang National Laboratory for Materials Science,
Institute of Metal Research, Chinese Academy of Science, School of
Materials Science and Engineering, University of Science and
Technology of China, 110016, Shenyang, China}
\affiliation{University of Chinese Academy of Sciences, Beijing,
100049, China}

\author{Jiangxu Li}
\affiliation{Shenyang National Laboratory for Materials Science,
Institute of Metal Research, Chinese Academy of Science, School of
Materials Science and Engineering, University of Science and
Technology of China, 110016, Shenyang, China}

\author{Ronghan Li}
\affiliation{Shenyang National Laboratory for Materials Science,
Institute of Metal Research, Chinese Academy of Science, School of
Materials Science and Engineering, University of Science and
Technology of China, 110016, Shenyang, China}

\author{Qing Xie}
\affiliation{Shenyang National Laboratory for Materials Science,
Institute of Metal Research, Chinese Academy of Science, School of
Materials Science and Engineering, University of Science and
Technology of China, 110016, Shenyang, China}
\affiliation{University of Chinese Academy of Sciences, Beijing,
100049, China}

\author{Hui Ma}
\affiliation{Shenyang National Laboratory for Materials Science,
Institute of Metal Research, Chinese Academy of Science, School of
Materials Science and Engineering, University of Science and
Technology of China, 110016, Shenyang, China}

\author{Dianzhong Li}
\affiliation{Shenyang National Laboratory for Materials Science,
Institute of Metal Research, Chinese Academy of Science, School of
Materials Science and Engineering, University of Science and
Technology of China, 110016, Shenyang, China}

\author{Yiyi Li}
\affiliation{Shenyang National Laboratory for Materials Science,
Institute of Metal Research, Chinese Academy of Science, School of
Materials Science and Engineering, University of Science and
Technology of China, 110016, Shenyang, China}

\author{Xing-Qiu Chen}
\email[Corresponding author:]{xingqiu.chen@imr.ac.cn}
\affiliation{Shenyang National Laboratory for Materials Science,
Institute of Metal Research, Chinese Academy of Science, School of
Materials Science and Engineering, University of Science and
Technology of China, 110016, Shenyang, China}

\date{\today}

\begin{abstract}
Topological semimetals with several types of three-dimensional (3D)
fermion of electrons, such as Dirac fermions, Weyl fermions, Dirac
nodal lines and triply degenerate nodal points have been
theoretically predicted and then experimentally discovered in the
electronic structures of a series of solid crystals. In analogy of
various typical fermions, topological mechanical states with two
type of bosons, Dirac and Weyl bosons, were also experimentally
reported in some macroscopic systems of kHz frequency and with a
type of doubly-Weyl phonons in atomic vibrational framework of THz
frequency of solid crystal was also recently predicted. However, to
date no triply degenerate nodal point of phonon beyond the
conventional Dirac, Weyl and doubly-Weyl phonons has been reported.
Here, through first-principles calculations, we have reported on the
prediction that the WC-type TiS, ZrSe, and HfTe commonly host the
unique triply degenerate nodal point of phonon in THz frequency due
to the occurrence of the phonon band inversion between the doubly
degenerate planar vibrational mode and the singlet vertical
vibrational mode at the boundary A point of the bulk Brillouin zone.
Quasiparticle excitations near this triply degenerate nodal point of
phonons are three-component bosons, different from the known
classifications. The underlying mechanism can be attributed to the
leading role of the comparable atomic masses of constituent elements
in compounds in competition with the interatomic interaction.
Additionally, the electronic structures in their bulk crystals
exhibit the coexisted triply degenerate nodal point and Weyl
fermions. The novel coexistence of three-component bosons,
three-component fermions and Weyl fermions in these materials thus
suggest an enriched platform for studying the interplay between
different types of fermions and bosons.
\end{abstract}

\maketitle

Topological semimetals \cite{H.Weng2016,C.K2016,A.Bansil2016,S.Rao}
are one of the fast growing families in the frontier of material
sciences and condensed matter physics due to their unique density of
states, transport properties and novel topological surface states as
well as potential applications for use in quantum computers and
spintronics. To date, it is well-known that topological semimetals
highlight several main types of interesting fermions in crystal
solids, such as three-dimensional (3D) Dirac cones
\cite{Z.Wang2012,Cheng.X2014,Young2012,Liu.Z2014,Xu2015,
S.M2012,Z.W2013,Z.K2014,Neupane2014,Du.Y2015,J.Hul}, Weyl
nodes\cite{S.Murakami2007,X.Wan2011,G.Xu2011,S.Y2015,Shekhar2015,S.Y_02015,
H.Weng2015,S.M2014,L.Lu2015,B.Q2015,B.Q_02015,S.-Y2015,
L.Yang2015,Y.Zhang2016,A.A2015,Xu
SY2016,Chang2016,Yang2016,Singh2012,Ruan2016}, Dirac nodal
lines\cite{Fang.C2016,Ryu2002,Heikkila2011,Burkov.A2011,Ronghan2016,
Weng.H.M2015,Yu.R2015,Kim2015,Xie.L2015,M.G.Zeng2015,Lu.L2013,Mullen2015},
triply degenerate nodal
points\cite{B.Bradlyn2016,G.W2016,H.Weng_02016,H.Weng_12016,
Zhu2016,Lv2016,G.Chang2016,He.J2017,Ding.H2017}, and even
beyond\cite{B.Bradlyn2016}. In addition, their realization in
crystal solids also renders important because they are providing the
ways to study elementary particles, which were long-sought and
predicted ones, in high-energy physics. Importantly, in similarity
to various fermions of electrons, the exciting progresses of the
bosons (vibrational phonons) have been also predicted
\cite{Lu.L2013} or observed in the 3D momentum space of solid
crystals with the topological vibrational states, such as Dirac,
Weyl and line-node phonons in photonic crystals only with
macroscopic systems of kHz frequency
\cite{Lu.L2013,Huber.S2016,Prodan.E2009,Chen.B2014,Yang.Z2015,
Wang.P2015,Xiao.M2015,Nash.L2015,Susstrunk2015,
Mousavi2015,Fleury2016,Rocklin2016,He2016,Lu2016,Susstrunk2016} and,
even most recently, theoretically predicted doubly-Weyl phonons in
transition-metal monosilicides with atomic vibrations at THz
frequency \cite{Zhang.T2017}. However, to date no three-component
bosons has been reported, although three-component fermions have
been experimentally discovered in the most recent work of MoP
\cite{Ding.H2017}.

The three-component bosons would possibly occur in atomic solid
crystals because three-fold degeneracy can be protected by lattice
symmetries, such as symmorphic rotation combined with mirror
symetries and non-symmorphic symmetries, as what was already
demonstrated to be triply degenerated points of electronic fermions
(three-component fermions) in the solid crystals
\cite{B.Bradlyn2016,G.W2016,H.Weng_02016,H.Weng_12016,Zhu2016,Lv2016,
G.Chang2016,He.J2017,Ding.H2017}. In addition to the importance of
seeking the new type of three-component boson, the topological
phonon states in atomic THz frequency will be extremely interesting,
because they could certainly enable materials to exhibit novel heat
transfer, phonon scattering and electron-phonon interactions, as
well as other properties related with vibrational modes, such as
thermodynamics. Within this context, through first-principles
calculations we report on the triply degenerate nodal points of
topological phonons in three compounds of TiS, ZrSe and HfTe.
Interestingly, these three materials simultaneously exhibit the
novel coexistence of three-component bosons, three-component
fermions, and two-component Weyl fermions, which provide attractive
candidates to study the  interplays between topological phonons and
two types of topological fermions in the same solid crystals.

Recently, the type of WC-type materials (Fig. \ref{fig1}(a)),
including ZrTe, TaN, MoP and WC has been theoretically reported to
host the coexistence of triply degenerate nodal points (TDNPs) of
electronic fermion and Weyl points (WPs) around the Fermi level.
Interestingly, this type of coexisted fermions of TDNPs and WPs have
been recently confirmed in MoP
\cite{H.Weng_02016,H.Weng_12016,Zhu2016,
Lv2016,G.Chang2016,He.J2017,Ding.H2017}. Within this context, here
we further extend this family by proposing eight compounds (TiS,
TiSe, TiTe, ZrS, ZrSe, HfS, HfSe and HfTe), which are isoelectronic
and isostructural to ZrTe. Among these eight compounds, four
compounds of TiS, ZrS, ZrSe$_{0.90}$, and Hf$_{0.92}$Se as well as
ZrTe were experimentally reported to have the same WC-type structure
\cite{Hahn_01959,Harry1957,Steiger1970,Hahn1959,Schewe1994,Sodeck1979,G.O2001,G.O2014}.
Because of no any experimental data available for the remaining four
compounds of TiSe, TiTe, HfS, and HfTe, here we have assumed that
they also crystallizes in the same WC-type structure. For five
experimentally known compounds TiS, ZrS, ZrSe, ZrTe and HfSe, our
DFT calculations yield the good agreement of their equilibrium
lattice parameters with the experimental data as shown in Table S1
(supplementary materials). Their enthalpies of formation are derived
in supplementary Table S1, indicating their stabilities in the
thermodynamics and their phonon dispersions have no any imaginary
frequencies, revealing the stabilities in the atomic mechanical
vibrations.

\begin{figure}
\includegraphics[height=0.45\textwidth]{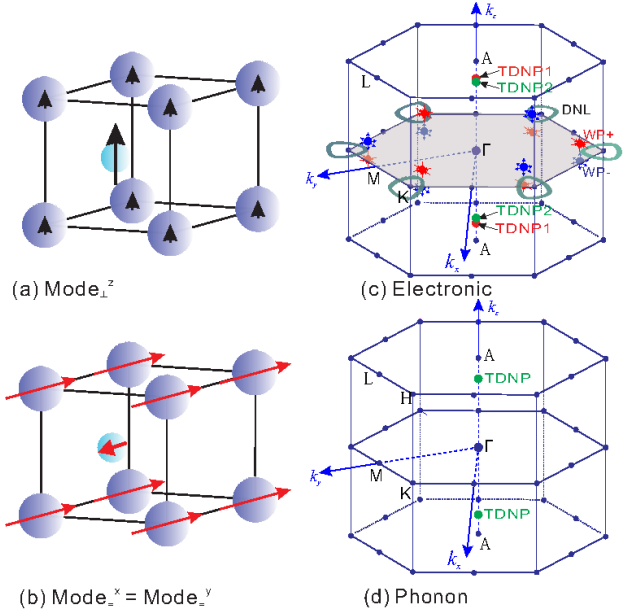}
\caption{WC-type crystal structure of \emph{MX} (\emph{M} = Ti, Zr,
Hf; \emph{X} = S, Se, Te), crystallizing in the simple hexagonal
crystal structure with the space group of $P\bar{6}m2$ (No. 187).
$M$ occupies the 1$a$ Wyckoff site (0, 0, 0) and $X$ locates at the
1$d$ (1/3, 2/3, 1/2) site. Panel (a) shows the phonon vertical
vibrational mode (Mode$_\perp^z$) along the $k_z$ direction at the
boundary (the high-symmetry A (0, 0, $\pi$/2) point) of the
Brilliouin zone (BZ). Panel (b) denotes the phonon planar
vibrational mode (Mode$_{=}^{x}$) along the $k_x$ direction, which
indeed is two-fold degenerate (called Mode$_{=}^{x,y}$ =
Mode$_{=}^{x}$ = Mode$_{=}^{y}$) because of its hexagonal symmetry.
Panel (c) The Brilliouin zone (BZ) in which the closed loops around
each $K$ point denotes the Dirac nodal lines (DNLs) of electrons
around the Fermi level when SOC is ignored. With the SOC inclusion
each DNL is broken into two Weyl points with the opposite chirality,
marked as blue (WP-) and red (WP+) balls and they coexist with the
triply degenerate nodal point (TDNP) of electronic structure
(namely, three-component fermion). Panel (d) shows the triply
degenerate nodal point (TDNP) of phonon dispersions (namely,
three-component boson) along the $\Gamma$-A direction.} \label{fig1}
\end{figure}

\begin{figure}
\includegraphics[height=0.48\textwidth]{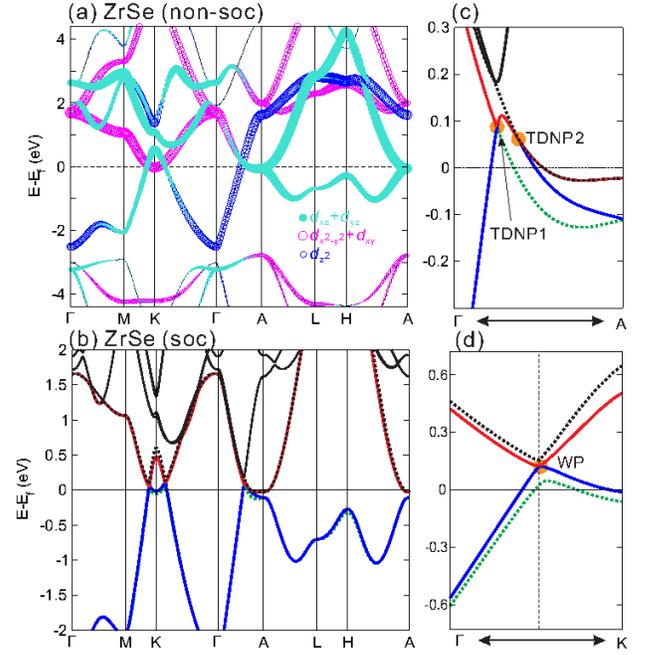}
\caption{Electronic structures of ZrSe. Panel (a): band structure
without the SOC inclusion shows (i) the band inversion between
$d_{xz}$+$d_{yz}$ and $d_{x^2-y^2}$+d$_{xy}$ at K and (ii) the band
inversion between $d_z^2$ and $d_{xz}$+$d_{yz}$ orbitals at A. Panel
(b): Electronic band structure with the SOC inclusion. Panel (c)
show the zoom-in visualization of two TDNP1 and TDNP2 along the
$\Gamma$-A direction in panel (b), whereas Panel (d) shows the
zoom-in bands crossing one WP (0.27314, 0.27314, $\pm$0.01628)
around the Fermi level with the SOC inclusion.} \label{fig1a}
\end{figure}

\begin{figure*}[!htp]
\vspace{0.5cm}
\includegraphics[height=0.37\textwidth]{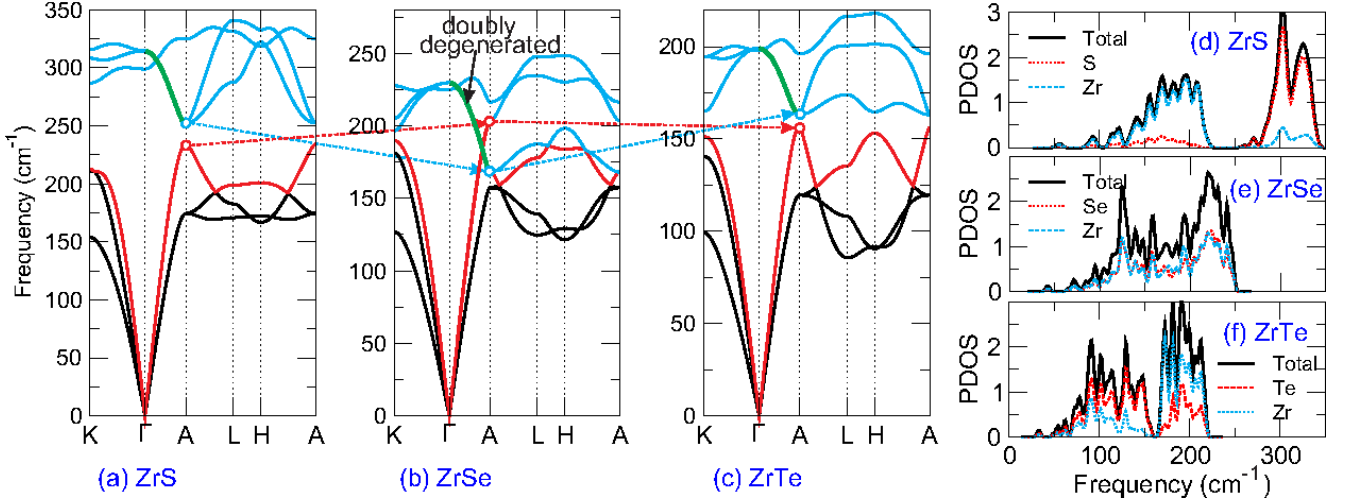}
\caption{Panels (a, b, and c): DFT-derived phonon dispersions of
ZrS, ZrSe and ZrTe, respectively. Panels (d, e, and f): DFT-derived
total and partial phonon densities of states (PDOS) of ZrS, ZrSe,
and ZrTe, respectively.} \label{fig2}
\end{figure*}

We have first elucidated the electronic band structures of these
nine compounds. Interestingly, their electronic structures are in
similarity to the case of ZrTe in Ref. \onlinecite{H.Weng_12016}. We
have selected ZrSe as a prototypical example to show the crucial
feature of electronic structures (details refer to Fig. S1-S3 in
supplementary materials). Without the spin-orbit coupling (SOC)
effect the two main features can be observed: In the first, a Dirac
nodal line (DNL as marked in Fig. \ref{fig1}(c)) centered at each
$K$ point in the $K_z$ = 0 plane is formed around the Fermi level
due to the linear crossing of the band inverted bands between the Zr
$d_{xz}$+$d_{yz}$ orbitals and Zr d$_{x^2-y^2}$+d$_{xy}$ orbitals
(Fig. \ref{fig1a}(a)). In the second, a six-fold degenerate nodal
point (Fig. \ref{fig1a}(a)) locating at (0, 0, 0.3025) along the
$\Gamma$-A direction around the Fermi level due to another band
inversion between the doubly degenerate Zr $d_{xz}$+$d_{yz}$ and Zr
$d_{z^2}$-like orbitals at the A point of the BZ. Because the masses
of both Zr and Se are not so slight that their SOC effects can not
be ignored. With the SOC inclusion, the derived electronic band
structure clearly exhibits apparent changes around the Fermi level:
Firstly, due to the lacking of inversion symmetry the spin splitting
bands appears and each DNL around the K point is indeed broken into
two Weyl points (WPs) with the opposite chirality (see WP+ and WP-
as marked in Fig. \ref{fig1}(c)). In total, there are six pairs of
WPs locating at both $k_z$ = $\pm$0.01628 plane slightly above and
below the $k_z$ = 0 planes. Interestingly, all these twelve WPs have
the same energy level (Fig. \ref{fig1a}(d) and supplementary Fig.
S1). Secondly, the SOC inclusion splits the six-fold generated nodal
point into the two triply degenerate nodal points (TDNP1 (0, 0,
0.2904) and TDNP2 (0, 0, 0.3146) as marked in Fig. \ref{fig1}(c) and
Fig. \ref{fig1a}(c)) along the $\Gamma$ to $A$ direction. Their
appearance is protected by the C$_{3z}$ rotation and the vertical
mirror symmetry, being the same as both ZrTe and TaN cases have
\cite{H.Weng_02016,H.Weng_12016}. Their non-trivial topological
property can be identified using the Wilson loop method \cite{add1}
(supplementary Fig. S1). Furthermore, we have derived the
topological non-trivial (0001) and (10$\bar{1}$0) surface states in
supplementary Figs. S2 and S3, showing the very similar non-trivial
surface states to ZrTe \cite{H.Weng_12016}. It still needs to be
emphasized that the other seven members in this family exhibit the
similar electronic structures featured with the coexisted TDNPs and
WPs in their electronic structures of bulk phases. Because of the
highly weak SOC effect in TiS, it is highly unique with the
coexisted six DNLs and two six-fold degenerate nodal points in its
electronic structure, which is exactly what happened for other eight
compounds when the SOC effect is ignored.

Besides their coexisted TDNPs and WPs in their electronic
structures, we have further found that the presence of the triply
degenerate nodal points (TDNPs) of phonon in three compounds of TiS,
ZrSe and HfTe after a systemical analysis of their phonon
dispersions (supplementary Fig. S4). Because each primitive cell
contains two atoms (Fig. \ref{fig1}(a)), its phonon dispersion has
six vibrational branches consisting of three acoustic and optical
ones, respectively. As compared with the computed phonon dispersions
in Fig. \ref{fig2}(a, b, and c) and their phonon densities of states
in Fig. \ref{fig2}(d, e, and f) of the isoelectronic ZrS, ZrSe and
ZrTe compounds, it can be clearly seen that a well-separated
acoustic-optical gap can be clearly observed in both ZrS and ZrTe
with their smallest direct gap at the A point (0, 0, $\pi$/2) on the
boundary of the BZ. The specified analysis uncovered that for both
ZrS and ZrTe compounds the top phonon band of the gap at the A point
is comprised with the doubly degenerate vibrational mode of phonons
in which both Zr and S (or Te) atoms, oppositely and collinearly,
displaces along either $x$ or $y$ direction (Mode$_{=}^{x,y}$ as
marked in Fig. \ref{fig1}(b)). The vibrational amplitude of the
Mode$_{=}^{x,y}$ are contributed nearly 100\% by the Zr atom, rather
than S (or Te) atoms. The bottom phonon band of the gap at the A
point is a singlet state originated from the vertical vibration mode
at which both Zr and S (or Te) atoms collinearly move in the same
$k_z$ direction (Mode$_{\perp}^{z}$ as marked in Fig.
\ref{fig1}(a)). But its amplitude of this Mode$_{\perp}^{z}$ are
almost fully dominated by the displacement of S (or Te) atoms.

In contrast to both ZrS and ZrTe in Fig. \ref{fig2}, the case of
ZrSe shows no acoustic-optical gap (Fig. \ref{fig2}(b)), as
illustrated by its phonon density of states in \ref{fig2}(e)). It
has been noted that the planar Mode$_{=}^{x,y}$ at the A point
becomes now lower in frequency than the vertical Mode$_{\perp}^{z}$.
Accordingly, this fact corresponds to the occurrence of so-called
phonon band inversion at the A point, indicating its non-trivial
topological mechanical property in ZrSe. It means the unusual fact
that around A point the optical phonon bands inverts below the
acoustic band which normally should have a lower frequency.
Physically, within the (quasi)harmonic approximation the vibrational
frequency, $\omega$, have to be proportional to $\sqrt{\beta/m}$ at
the boundary of the BZ. Here, $\beta$ is the second-order force
constant - the second derivative of the energy following a given
vibrational mode as a function of the displacement and $m$ the
atomic mass. Therefore, as seen in Fig. \ref{fig2}(b) for ZrSe the
occurrence of the phonon band inversion at the boundary A point is
certainly induced by both $\beta$ and $m$ which are both determined
by the planar Mode$_{=}^{x,y}$ and the vertical Mode$_{\perp}^{z}$
at the A point. Following this consideration, we have defined the
dimensionless ratio $\tau$ as follows,
\begin{equation}
\tau = \frac{\sqrt{\beta_=/m_=}}{\sqrt{\beta_\perp/m_\perp}},
\end{equation}
where $\tau$ indeed specifies the comparison between the frequencies
of both Mode$_{=}^{x,y}$ and Mode$_{\perp}^z$. With $\tau$ $>$ 1 the
material shows no band inversion, thereby being a trivial
vibrational mechanical state, whereas $\tau$ $<$ 1 implies the
topological non-trivial mechanical states exist due to the
appearance of the phonon band inversion. If $\tau$ $\approx$ 1, the
targeted material would locate at the phase boundary between trivial
and non-trivial mechanical states. With such a definition, we
further plot the $\beta$ in Fig. \ref{fig3}(a) in the sequence of
ZrS, ZrSe and ZrTe. It has been found that, only with the
second-order force constants of $\beta_{=}$ and $\beta_{\perp}$
(Fig. \ref{fig3}(a)) it is not enough to induce the phonon band
inversion. This fact is in agreement with the Equ. (1) although the
$\beta_{=}$-$\beta_{\perp}$ difference is the smallest in ZrSe among
these three cases in Fig. \ref{fig3}(a). In further combination with
their atomic masses, the topological non-trivial mechanical property
occurs for ZrSe because its $\tau$ value is now 0.68, smaller than
1. Similar analysis has been performed for the trivial cases of both
ZrS and ZrTe which all have a $\tau$ value to be larger than 1, in
agreement with no phonon band inversion. Furthermore, for all nine
compounds in this family we compiled their $\tau$ values as a
function of the ratio ($\delta$) of the atomic masses related with
Mode$_{=}^{x,y}$ over Mode$_{\perp}^z$ (namely, $\delta$ =
$m$(Mode$_{=}^{x,y}$)/$m$(Mode$_{\perp}^{z}$)) in Fig.
\ref{fig3}(b). With increasing the ratio of the atomic masses, the
$\tau$ value increases in a nearly linearity. This implies that if
atomic masses in a targeted material highly differ its possibility
to become topological non-trivial mechanical property is extremely
low. However, if they have the comparable atomic mass with the
$\delta$ ratio close to 1 the possibility with the non-trivial
topological property is high. Following this model, we have further
uncovered that, because the $\tau$ value is smaller than 1, both TiS
and HfTe have topological non-trivial mechanical property as what
ZrSe does (Fig. \ref{fig3}(b)). The findings for both TiS and HfTe
are in accordance with the DFT-derived phonon dispersions in
supplementary Fig. S4. However, the mechanical properties of other
members in this family are trivial. These facts imply that in these
materials the difference between the atomic masses of constituents
in compound play a key role in inducing the phonon band inversion,
as seen for three non-trivial cases of TiS, ZrSe and HfTe whose
$\delta$ value are all around 1.

\begin{figure}[!htp]
\includegraphics[height=0.30\textwidth]{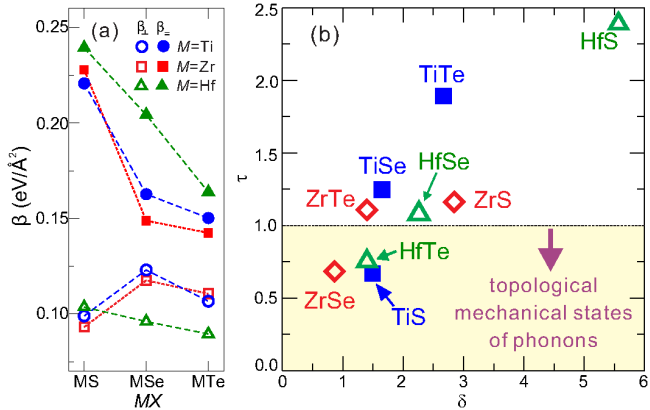}
\caption{Panel (a): DFT-derived second-order force constant at the A
point for both the two-fold degenerate planar vibrational
Mode$_=^{x,y}$ and the vertical vibrational Mode$_{\perp}^z$. Panel
(b): The derived parameter $\tau$ from Equ. (1) as a function of the
$\delta$ value, as defined in the main text, for all nine
compounds.} \label{fig3}
\end{figure}

\begin{figure}[!htp]
\vspace{0.2cm}
\includegraphics[height=0.45\textwidth]{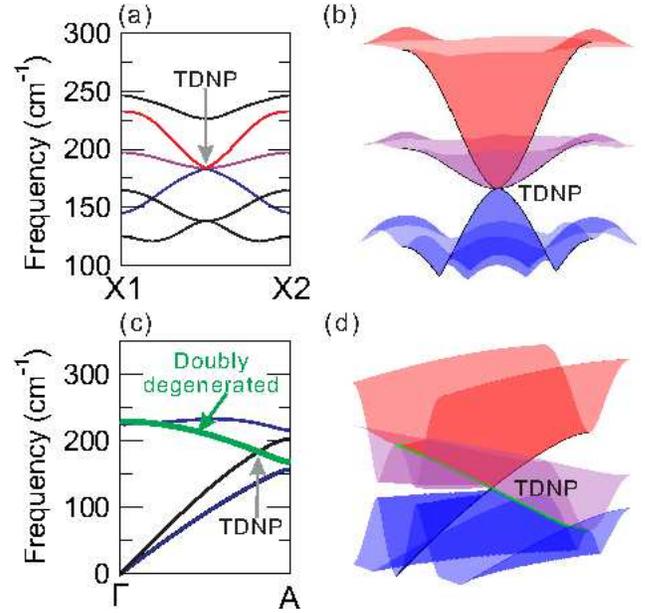}
\caption{Panels (a and b): DFT-derived phonon dispersions to
elucidate phonon TDNPs of ZrSe along the X1 (-$\pi$/2, 0, 0) to X2
($\pi$/2, 0, 0) and $\Gamma$-A directions, respectively. Panels (c
and d): Zoom-in 3D visualization of phonon TDNPs in the $k_z$ = 0
and $k_y$ = 0 planes, respectively.} \label{fig4}
\end{figure}

Importantly, as accompanying with the occurrence of the phonon band
inversion the TDNPs, featured by a linear crossing of the
frequencies between the acoustic and optical bands, unavoidably
appears at (0, 0, $k_z$ = $\pm$0.40769) along the $\Gamma$-A
direction in the BZ (Fig. \ref{fig2}(b)) for ZrSe. It implies that
at this phonon TDNP the planar Mode$_{=}^{x,y}$ and the vertical
Mode$_{\perp}^Z$ at (0, 0, $k_z$ = $\pm$0.40769) strictly have the
same frequency of 183.9 cm$^{-1}$. The TDNP locates at (0, 0, $k_z$
= $\pm$0.40382) with 293.4 cm $^{-1}$ for TiS and at (0, 0, $k_z$ =
$\pm$0.43045) with 133.3 cm $^{-1}$ for HfTe. To elucidate the 3D
TDNP shape of ZrSe, we also plotted the zoom-in dispersions within
both $k_{z}$ = 0 and $k_y$ = 0 planes of BZ in Fig. \ref{fig4}. From
both Fig. \ref{fig4}(a) and (c) in the $k_{z}$ = 0 the TDNP of
phonon can be clearly visualized to have an isotropic shape.
However, in the $k_y$ = 0 plane the phonon bands around the TDNP are
highly complex with the helicoid shape (Fig. \ref{fig4}(b) and (d)).
Therefore, the topological surface states related to this TDNP of
phonon may not be observed, since it indeed does not leave a bulk
gap when projected onto any open surface.

Summarizing, through first-principles calculations we have
discovered three WC-type materials of TiS, ZrSe and HfTe not only
host three-component bosons featured by triply degenerate nodal
points in its mechanical phonon dispersions but also exhibits
three-component fermions feature by triply degenerated nodal points
and six pairs of two-component Weyl fermions with opposite chirality
in its electronic structures of the bulk crystal. Surprisingly, the
novel coexistence of the main features of (\emph{i}) three-component
bosons, (\emph{ii}) three-component fermions, and (\emph{iii})
two-component Weyl fermions and, in particular, both three-component
bosons and three-component fermions locate at the nearly same
momentum (Fig. \ref{fig1}(c and d)) along the $\Gamma$-A direction
could couple to each other through electron-phonon interactions,
which would provide a wonderful platform to study the interplays
between different types of topological electron excitations and
topological phonons within the atomistic scale for potential
multi-functionality quantum-mechanical properties.

\vspace{0.5cm} \noindent {\bf Acknowledgments} We thank H. M. Weng
for valuable discussions. Work was supported by the ``Hundred
Talents Project'' of the Chinese Academy of Sciences and by the
National Natural Science Foundation of China (Grant Nos. 51671193
and 51474202) and by the Science Challenging Project No. TZ2016004.
All calculations have been performed on the high-performance
computational cluster in the Shenyang National University Science
and Technology Park and the National Supercomputing Center in
Guangzhou (TH-2 system).

\vspace{0.5cm} \noindent {\bf Methods}

Within the framework of the density functional theory (DFT) and the
density functional perturbation theory (DFPT), we have performed the
calculations on the structural optimization, the electronic band
structures, the phonon calculations and surface electronic band
structures. Both DFT and DFPT calculations have been performed by
employing the Vienna \emph{ab initio} Simulation Package (VASP)
\cite{G.Kresse1993,G.Kresse1994,G.Kresse1996, G.Kresse_01996}, with
the projector augmented wave (PAW) pseudopotens
\cite{P.E1994,G.Kresse1999} and the generalized gradient
approximation (GGA) within the Perdew-Burke-Ernzerhof (PBE)
exchange-correlation functional\cite{J.P1996}. The adopted PAW-PBE
pseudopotentials of all elements treat semi-core valence electrons
as valence electrons. A very accurate optimization of structural
parameters have been calculated by minimizing the interionic forces
below 0.0001 eV/\AA\,. The cut-off energy for the expansion of the
wave function into the plane waves was 500 eV. The Brillouin zone
integrations were performed on the Monkhorst-Pack k-meshes
(21$\times$21$\times$23) and were sampled with a resolution of
2$\pi$ $\times$ 0.014\AA\,$^{-1}$. The band structures, either with
or without the inclusion of spin-orbit coupling (SOC), have been
performed by the Gaussian smearing method with a width of smearing
at 0.01 eV. Furthermore, the tight-binding (TB) through Green's
function methodology \cite{M.P1985,Weng2014,Weng2015} were used to
investigate the surface states. We have calculated the Hamiltonian
of tight-binding (TB) approach through maximally-localized Wannier
functions (MLWFs) \cite{N.Marzari1997,I.Souza2001} by using the
Wannier 90code \cite{A.A2008}.

\clearpage

\maketitle

\vspace{0.5cm} \noindent {\bf \Large Supplementary Materials}

\begin{enumerate}
  \item \textbf{Table S1}: optimized lattice parameters of $MX$
  \item \textbf{Figure S1}: Evolution of the electronic structure around the Weyl points
  (WPs) in ZrSe
  \item \textbf{Figure S2}: Surface electronic band structures
  of (0001) and (10$\bar{1}$0) surfaces of ZrSe
  \item \textbf{Figure S3}: Fermi surfaces of the (0001) and (10$\bar{1}$0) surfaces of
  ZrSe
  \item \textbf{Figure S4}: DFT-derived phonon dispersions of
  the nine $MX$ compounds
\end{enumerate}

\subsection{Supplementary Table S1}

We have optimized the lattice structures of nine $MX$ compounds with
the WC-type structure. Table S1 summarizes all optimized lattice
constants as compared with the available experimental data. Among
these nine compounds, five compounds of TiS, ZrS, ZrSe$_{0.90}$, and
Hf$_{0.92}$Se as well as ZrTe were experimentally reported to have
the same WC-type structure
\cite{Hahn_01959,Harry1957,Steiger1970,Hahn1959,Schewe1994,Sodeck1979,G.O2001,G.O2014}.
Because of no any experimental data available for the remaining four
compounds of TiSe, TiTe, HfS, and HfTe, here we have assumed that
they also crystallizes in the same WC-type structure. For five
experimentally known compounds TiS, ZrS, ZrSe, ZrTe and HfSe, our
DFT calculations yield the good agreement of their equilibrium
lattice parameters with the experimental data as shown in Table S1.
Furthermore, their enthalpies of formation are derived in Table S1,
indicating their stabilities in the thermodynamics.

\subsection{Electronic band structures}

To elucidate the electronic band structure of these compounds, we
have first repeated the calculations of ZrTe and obtained the
electronic band structures are very similar to the reported data in
Ref. \onlinecite{H.Weng_12016}, indicating the reliability of our
current calculations. Remarkably, the derived electronic band
structures of other compounds in this family are all similar to that
of ZrTe.

The evolution of the derived electronic band structures around one
of WPs for ZrSe is illustrated in supplementary Fig. S1(a,b,c) and
it can be clearly seen that the WP appears around the Fermi level in
supplementary Fig. S1c. The topological invariant can be identified
using the so-called Wilson loop method \cite{add1}. As shown in
supplementary Fig. S1(d and e), the calculated evolution of the
Wannier centers formed along the k$_y$ direction in the two k$_z$ =
0 and $\pi$ planes. It can be seen that the Z$_2$ numbers (namely,
counting the times of Wannier center crosses a reference line) of
both these planes are odd, indicating their topological non-trivial
feature.

\begin{table}[!t]
\begin{center}
\caption{(\textbf{Supplementary Table S1)}: DFT-derived lattice
constants $a$ (\AA) and $c$ (\AA) and enthalpy of formation
(eV/atom) of single crystals, in comparison with available
experimental data. \label{tab1}}
\begin{ruledtabular}
\begin{tabular}{ccccc}
                & \emph{a}           & \emph{c}     & $ \Delta$H   & \\
\hline
 TiS  &  3.287    &  3.210            &               &  Expt. Ref. \onlinecite{Hahn_01959}  \\
      &  3.267    &  3.223            &    -1.50   & Calc. \\
 \hline
 TiSe & 3.419    &  3.402                       &    -1.28                  & Calc.   \\
 \hline
 TiTe & 3.669    &  3.656                      &    -0.64                 & Calc.  \\
 \hline
 ZrS  & 3.446    &  3.445                           &                      &    Exp. Ref. \onlinecite{Harry1957,Steiger1970}   \\
      & 3.460    &  3.475                     &    -1.65                  & Calc.  \\
 \hline
 ZrSe$_{0.90}$   & 3.551   &  3.615                           &                      &   Exp. Ref. \onlinecite{Hahn1959}   \\
 ZrSe            & 3.584   &  3.649                    &    -1.49                 & Calc.  \\
 \hline
 ZrTe            &  3.760 & 3.860 & & Exp. Ref.
 \onlinecite{Sodeck1979,G.O2001,G.O2014} \\
                 & 3.798 & 3.895 & -0.91 & Calc.\\
 \hline
 HfS             & 3.395    &  3.447                    &    -1.54                 & Calc.   \\
 \hline
 Hf$_{0.92}$Se   &  3.4958    &  3.6069                           &                      &   Exp. Ref. \onlinecite{Schewe1994}   \\
                 & 3.5173    &  3.6365                   &    -1.32               & Calc.   \\
 \hline
 HfTe            & 3.739    &  3.885                   &    -0.68               & Cal.   \\
\end{tabular}
\end{ruledtabular}
\end{center}
\end{table}

\begin{figure}
\includegraphics[height=0.35\textwidth]{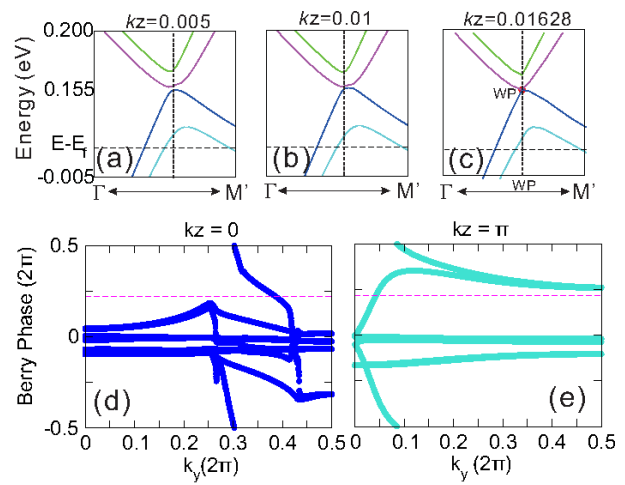}
\caption{(\textbf{Supplementary Figure S1}) Electronic band
structures around the WP node at $K_z$ = 0.005 in panel (a), 0.010
in panel (b), and 0.01628 (exactly corresponding to the WP node) in
panel (c), respectively. Panels (d and e) denote the derived Wilson
loops of ZrSe which show the $k_y$ evaluation of the Berry phases of
all occupied bands along the $k_x$ direction in both $k_z$ = 0 and
$k_z$ = $\pi$ planes, respectively. } \label{fig3s}
\end{figure}

To further clarify the topological feature in ZrSe, we have
calculated the surface electronic structures on its (0001) and
(10$\bar{1}$0) surfaces (see supplementary Fig. S2), clearly
indicating the topological surface states. In addition, we also plot
the Fermi surface of both surfaces in supplementary Fig. S3. On the
(0001) Fermi surface the two TDNPs along $\Gamma$ to $A$ direction
in its bulk phase are both projected onto the $\bar{\Gamma}$ point,
becoming invisible due to the overlapping with the projection of the
bulk electronic bands. Each pair of WPs above and below the $k_z$ =
0 plane which have different chirality in the bulk phase will be
projected onto the same point, totally forming six projected nodes.
These six projected nodes on the (0001) surface are further
connected by Fermi arcs, resulting in the appearance of two
triangle-like loops, as illustrated in supplementary Fig. S3(a). It
is impossible to see these six projected nodes in supplementary Fig.
S3(a) because their energy is 155 meV above the Fermi level. By
changing the chemical potential to 155 meV, the Fermi surface gives
rise to the clear visualization of six projected nodes in
supplementary Fig. S3(c). In order to visualize the Weyl nodes, we
further calculated (10$\bar{1}$0) Fermi surface at the energy level
of 155 meV above the Fermi level in supplementary Fig. S3(d). On
this surface, the six pairs of WPs with opposite chirality are
projected to different positions. Two WPs with same chirality are
projected to the same point on the (10$\bar{1}$0) surface (called
WP1) and the projected points of other WPs are labeled as WP2. It
can be clearly seen that the projected Weyl points are connected by
Fermi arcs. For each WP1 point, there is one arc connecting it by
going through the $\bar{\Gamma}$ - $\bar{M}$ path, whereas, for each
WP2 there are two arcs connecting them in supplementary Fig. S3(d).
In addition, on both (0001) and (10$\bar{1}$0) surfaces it is
impossible to see TDNPs because the projection of TDNPs are all
overlapped with bulk electronic bands, as illustrated in Fig.
supplementary Fig. S3(c).

\begin{figure*}
\includegraphics[height=0.35\textwidth]{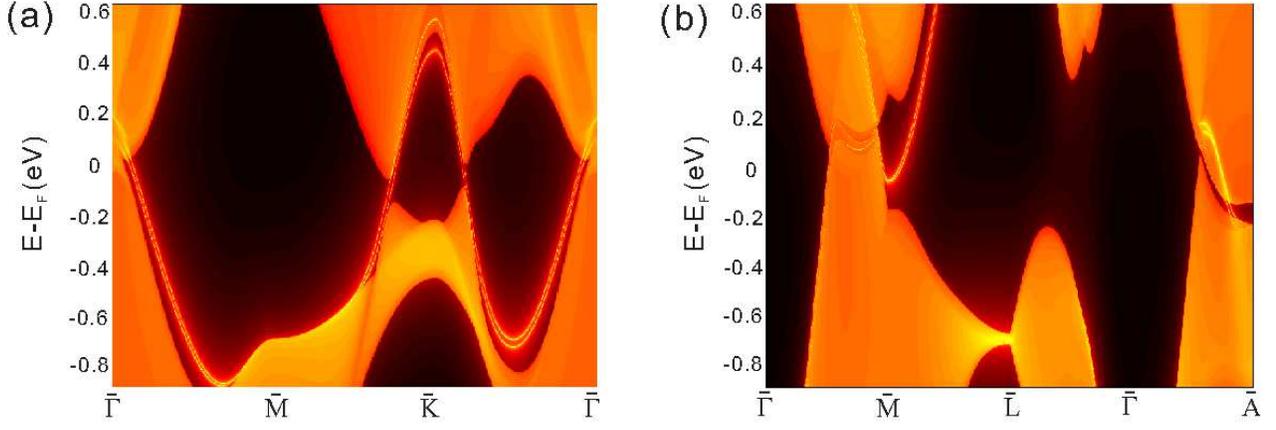}
\caption{(\textbf{Supplementary Figure S2}) Calculated surface
electron structures of both (0001) and (10$\bar{1}$0) surfaces of
ZrSe.} \label{fig4s}
\end{figure*}

\begin{figure*}
\includegraphics[height=0.80\textwidth]{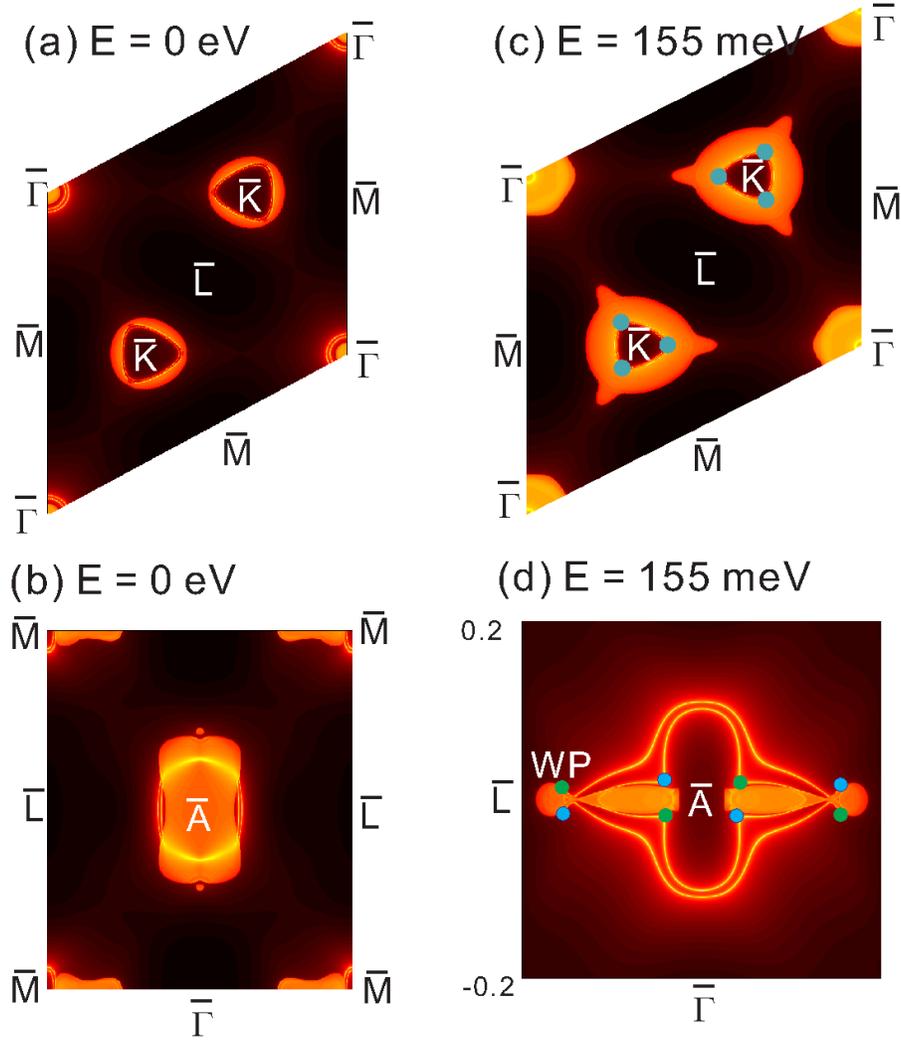}
\caption{(\textbf{Supplementary Figure S3}) The Fermi surfaces of
(0001) (panels: a and c) and (10$\bar{1}$0) (panels: b and d)
surfaces of ZrSe. Panels (a) and (b) denote the Fermi surfaces at
the Fermi level, while panels (c) and (d) refer to the Fermi
surfaces with a chemical potential of 155 meV above the Fermi level.
The solid circles indicate the projected WPs in the (10$\bar{1}$0)
surfaces.} \label{fig5s}
\end{figure*}

Through our calculations the other eight members in this family all
exhibit the similar electronic structures with the coexisted TDNPs
and WPs in their bulk phases, except for TiS. Because of the weak
SOC effect, TiS is highly unique with the coexisted six Dirac nodal
lines (DNLs) and two six-degenerated nodal points, which is exactly
what happened for ZrSe without the SOC effect.

\subsection{Phonon dispersions of MX}

Supplementary Figure S4 compiles the DFT-derived phonon dispersions
of all nine $MX$ compounds. Among them, only three compounds of TiS,
ZrSe and HfTe exhibit the non-trivial topological phonon states with
the appearance of the triply degenerate nodal points (TDNPs as
marked in supplementary Figure S4, which refers to three-component
bosons.

\begin{figure*}
\includegraphics[height=0.80\textwidth]{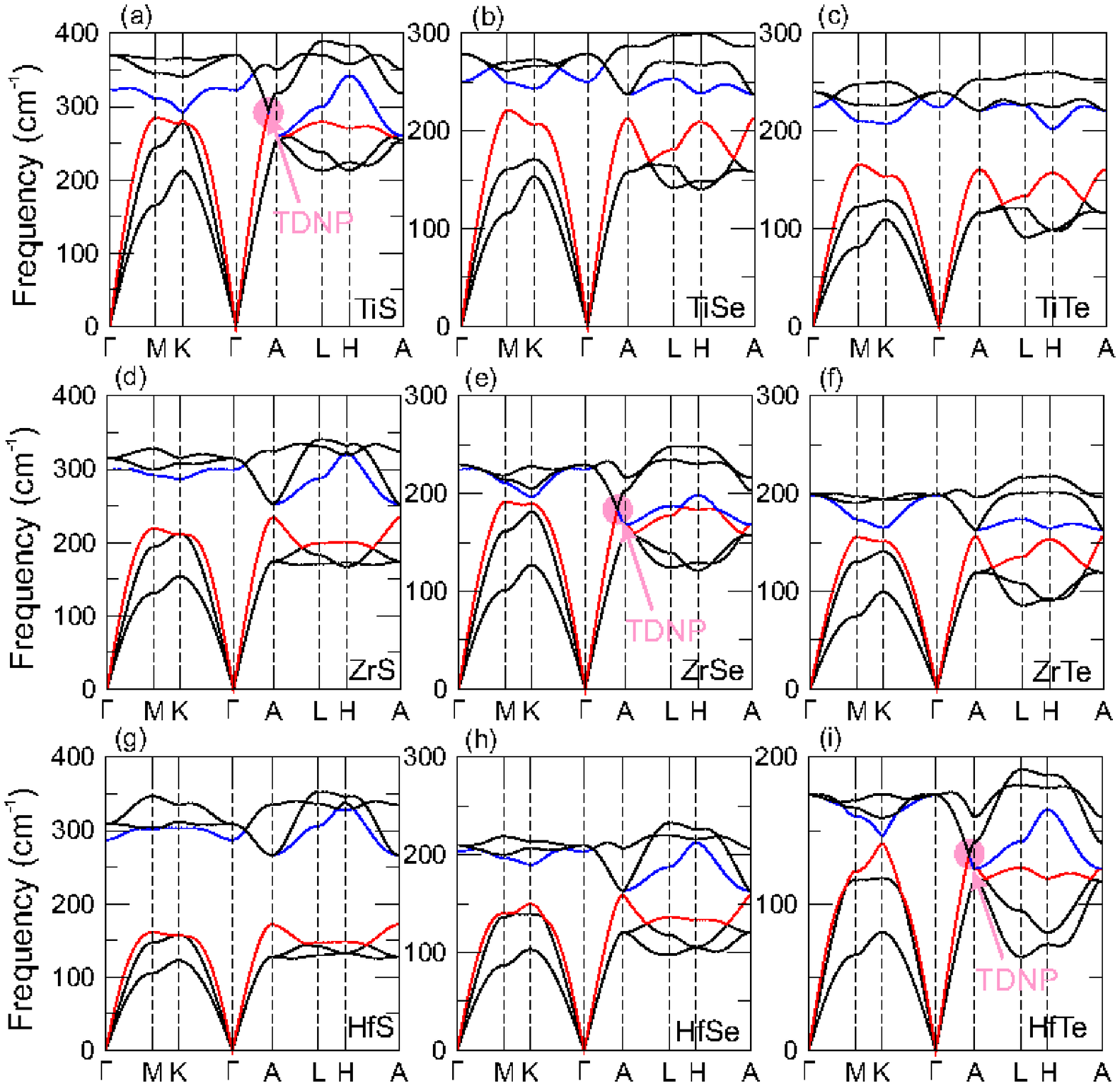}
\caption{(\textbf{Supplementary Figure S4}) DFT-derived phonon
dispersions of all nine \emph{MX} compounds with the WC-type
structure at their optimized equilibrium lattices.} \label{fig5s}
\end{figure*}
\end{document}